\documentclass[prl,floatfix]{revtex4}
\usepackage{graphics}
\usepackage{graphicx}

\newcommand{\eqref}[1]{(\ref{#1})}

\newcommand{\pd}[2]{{\frac{\partial #1}{\partial #2}}}

\newcommand{\be}{\begin{equation}}
\newcommand{\ee}[1]{\label{#1} \end{equation}}
\newcommand{\ba}{\begin{eqnarray}}
\newcommand{\ea}[1]{\label{#1} \end{eqnarray}}

\begin{document}

\title{ {\bf Entropy of expanding QCD matter}}

\author{Tam\'as S. Bir\'o
}
\affiliation{
 KFKI Research Institute for Particle and Nuclear Physics Budapest
}
\author{J\'ozsef Zim\'anyi
}
\affiliation{
 KFKI Research Institute for Particle and Nuclear Physics Budapest
}

\date{April 13. 2007}

\begin{abstract}
Using the lattice QCD equation of state for an isentropically
expanding fireball we follow the evolution of the effective
number of particles in an ideal gas, $N_{{\rm eff}}=pV/T$.
This number reduces to its third around the crossover temperature,
which helps to resolve the entropy problem inherent in 
hadronization models assuming quark coalescence.

\end{abstract}

\maketitle


\vspace*{7mm}


The hadronization of quark matter is most probably a complex, highly non-equilibrium
process. During a sudden cooling and expansion elementary particle correlations
are re-arranged while passing the crossover temperature of QCD.
It is therefore surprising that simple models, such as a nearly ideal quasiparticle
gas in quark matter and a hadronic resonance gas, are good fits to the
equation of state of the strongly interacting system studied in lattice QCD. 
Moreover during a sudden, quark-recombining
hadronization scenario the number of effective degrees of freedom is
dramatically reduced.

Hadronization models therefore either dynamically fragment quarks and gluons
to several hadrons, or face the problem how not to reduce the entropy. 
The simple quark-coalescence idea, based on the valence
quark model of hadrons, redistributes and bounds quasiparticles into hadrons
assuming an effectively low gluonic quasiparticle number in the formation process.
The question naturally arises whether a small quasiparticle number
can be realistically assumed while not reducing the entropy of the quark matter.
The argumentation pointing out an entropy reduction 
usually assumes a massless ideal gas full with gluons
on the QGP side and only the lightest hadrons (pions) on the hadron side.
This is not very realistic.
We show that the entropy obstacle can be circumvented by choosing appropriate
temperatures on the quark and hadron matter sides in accord with the equation
of state obtained from lattice QCD.

In this paper, supported by our recent analysis of lattice QCD equation of state (eos) 
data\cite{Biroetal}, we show that during an adiabatic expansion the number of
massive particles in a quasiparticle approach\cite{PESHIER} 
can be reduced to half or to one third in a realistic expansion time.
This quantitative property of the lattice QCD eos explains the success
of the massive quark matter coalescence hadronization model\cite{ALCOR,COALESCE,JPG}.

\begin{figure}

\includegraphics[width=0.45\textwidth,angle=-90]{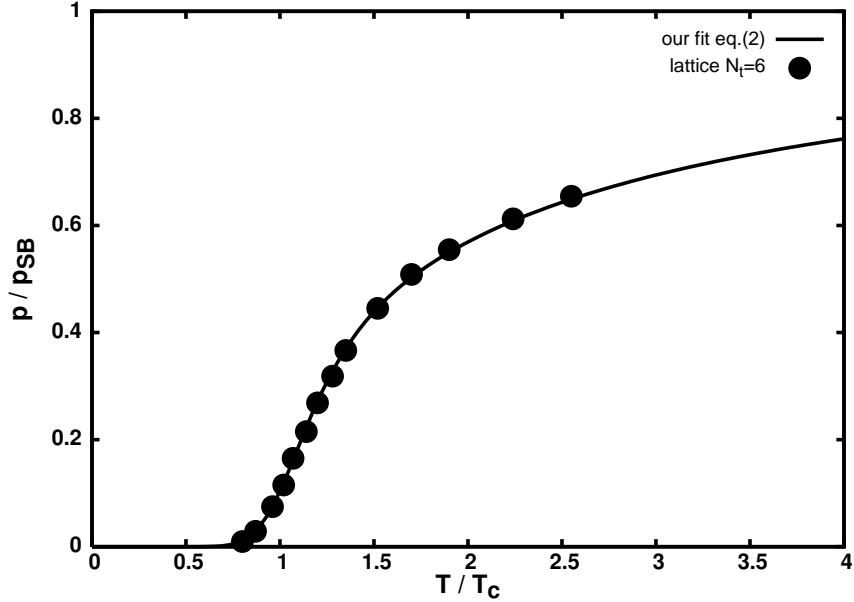}

\caption{ \label{FIG0}
 The normalized pressure as a function of the temperature fitted to
 lattice QCD eos data of Ref.\cite{FODOR}.
}
\end{figure}

The following fit to the continuum extrapolated pressure curve of Ref.\cite{FODOR} 
can be  considered (cf. Fig.\ref{FIG0})
\be
 p = c \: T^4 \: \sigma(T_c/T)
\ee{PRESSURE}
with $c=5.21$ and
\be
 \sigma(z) = \frac{1+e^{-a/b}}{1+e^{(z-a)/b}} \: e^{-\lambda z},
\ee{SIGMA}
where $z = T_c/T,  a=0.91, b=0.11$ and  $\lambda=1.08$. 
The Stefan-Boltzmann coefficient per degree of freedom, $\sigma(z)$, is
temperature dependent. This reflects the interaction, in particular mass generation,
in the lattice QCD eos\cite{Biroetal}. At infinite temperature the massless
ideal gas is expected, therefore $\sigma(0)=1$. Of course the data points of the QCD
lattice eos also can be used without any interpolating fit.

The number of particles in an ideal gas, which has the same pressure as
a given interacting system, can be defined through the thermodynamical potential. We refer to
this quantity as the effective number; it represents the number
of non- or weakly interacting quasiparticles:
\be
 N_{{\rm eff}} = \frac{1}{T} \, \Omega(V,T,\mu).
\ee{NEFF}
For homogeneous systems, as it is assumed by the extrapolation of lattice QCD
data to infinite volume, this potential is simply related to the pressure:
$\Omega(V,T,\mu)=V\: p(T,\mu)$.
The elementary particle number is on the other hand given by
\be
 N = \frac{\partial\Omega}{\partial\mu} = N_{{\rm eff}} +
 \left( \frac{\partial\Omega}{\partial\mu} - \frac{\Omega}{T}\right).
\ee{NELEM}
It is straightforward to see that the expression in the bracket vanishes and
hence $N_{{\rm eff}}=N$ in the Boltzmann approximation
\be
 \Omega(V,T,\mu) = \Omega_0(V,T) \, e^{\mu/T}.
\ee{BOLT_APP}
In this case, using homogeneity, the Boyle-Mariotte law is fulfilled:
\be
 pV = NT.
\ee{BOYLE-MARIOTTE}
For the quark coalescence model of hadronization the quasiparticle number,
$N_{{\rm eff}}$ is relevant. For a mixture of quasiparticle gases
\be
 \Omega = \sum_i \Omega_i(V,T) \, e^{c_i\mu_i/T},
\ee{ID_MIX}
and the effective number becomes
\be
 N_{{\rm eff}} = \frac{\Omega}{T} = \sum_i \frac{1}{c_i} N_i.
\ee{EFF_MIX}
It is a weighted sum of elementary particle numbers. In order to estimate the
hadron numbers by the coalescence principle, this quantity has to drop suitably
while the temperature drops and the volume expands. The main question to
clarify is whether keeping the entropy constant during the expansion an 
appreciable reduction in the quasiparticle number $N_{{\rm eff}}$ is possible
in accordance to the equation of state of an interacting quark-gluon plasma.

Using the definition of entropy, $S=\partial\Omega/\partial T$,
the effective number of equivalent non-interacting particles
(quasi quarks and gluons above $T_c$, hadrons below $T_c$) divided by the total entropy
can be extracted from the $p(T)$ curve as follows
\be
 \frac{N_{{\rm eff}}}{S} =  \frac{p}{T\pd{p}{T}}.
\ee{PARTICLE_NUMBER}
Substituting our fit for $\sigma(z)$, eq.(\ref{SIGMA}), we arrive at
\be
 {N_{{\rm eff}}} \: = \: S \, 
  \left( 4 + \frac{\lambda T_c}{T} + \frac{T_c}{bT\left(1 + e^{a/b-T_c/bT} \right)} \right)^{-1}.
\ee{NUMBER}

\begin{figure}

\includegraphics[width=0.45\textwidth,angle=-90]{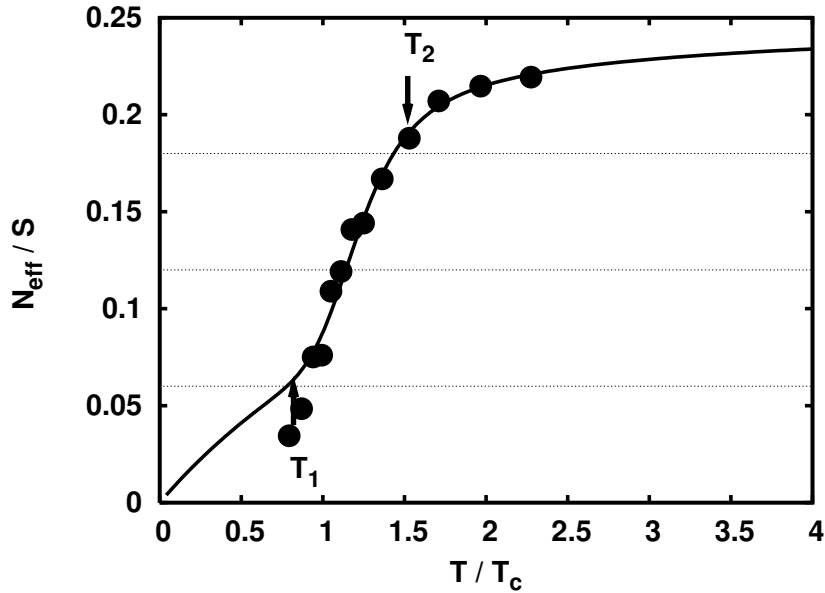}

\caption{\label{FIG1} 
 The effective number of particles divided by the total entropy
 for a non-interacting ideal gas, which reconstructs
 lattice QCD pressure as a function of temperature. A reduction 
 to $1/3$ is achieved by cooling from $T_2=254$ MeV to $T_1=137$ MeV.
}
\end{figure}

Fig.\ref{FIG1} plots the ratio $N_{{\rm eff}}/S$ as a function of the scaled temperature $T/T_c$
using the parameters $a=0.91, b=0.11, \lambda=1.08$. The big dots indicate the result
obtained from $p(T)$ data by numerical derivation.

The resulting curve in Fig.\ref{FIG1} shows two turning points: one below, one above
$T_c$. Considering a standard value, $T_c=167$ MeV, the turnings are located
at $T_1=0.82T_c=137$ MeV and $T_2=1.52T_c=254$ MeV. These points were estimated
as the endpoints of the steepest linear rise in the $N_{{\rm eff}}/S$ curve.
The effective number of ideal quasiparticles
falls to its one third between these two temperatures while keeping the total
entropy constant. This roughly corresponds to changing from a colored,
quark dominated gas to a color singlet one. The temperature $T_2$ is around
the color deconfinement temperature obtained earlier in pure gluon systems\cite{BIELEFELD}.

The fastest change in $N_{{\rm eff}}/S$ is concentrated to the temperature range between
$T_1$ and $T_2$. This may be considered as the transition region between the
high temperature QGP (where 
the quasiparticle masses are set by the scale $\lambda T_c = 180$ MeV) and
the hadronic phase near $T_c$  (dominated by heavy resonances\cite{HRESGAS}). The 
large number of different hadron types accommodates the high entropy
inherent in the QGP with liberated color degrees of freedom.


In an earlier work\cite{Biroetal} we have analyzed the lattice QCD equation
of state in terms of distributed mass ideal parton gas. We found that at temperatures
over $T_c$ the mass distribution starts at $\lambda T_c \approx 180$ MeV and has its maximum
around the valence quark mass, $m_v\approx 300 - 350$ MeV. Below $T_c$ the
characteristic minimal mass is about ten times higher, $10T_c \approx 1.67$ GeV.

We interpret Fig.\ref{FIG1} in the following way:
as the temperature decreases due to physical expansion of the volume
at constant entropy,  the number of quasiparticles drops by about a factor of three. 
We know from earlier studies\cite{BiroLVZ} that around the same temperature the
characteristic mass scale of the distributed mass quasiparticles can increase
by a factor of ten.
This reduction in the effective particle number definitely does not contradict to the
constant entropy. Thus we conclude that hadronization models utilizing such a
reduction in the number of particles do not necessarily reduce the total
entropy.


In order to connect this observation to the real time behavior of expanding quark
matter one has to explore the relation between the eos and the hydrodynamical
expansion of the perfect fluid. This expansion is isenthalpic, 
$dE+pdV=TdS+\sum_i \mu_i dN_i = 0$, if chemical transmutation rates
are neglected even isentropic $dS=0$. Since the QCD color deconfinement transition
is a crossover, no latent heat is released, and therefore the hadronization
is neither an exothermic nor an endothermic process. We consider $dS=0$.
A local volume is defined by the expansion rate
\be
 \partial^{\mu} u_{\mu} \: = \: \frac{1}{V} \frac{dV}{d\tau},
\ee{FOUR-DIV}
where $d/d\tau=u^{\mu}\partial_{\mu}$ is the comoving derivative.
The cooling takes place due to the reduction of the internal energy $E=eV$
by the work of the pressure $p$ on the local expansion: 
\be
 \frac{de}{d\tau} + (e+p)\frac{1}{V}\frac{dV}{d\tau} \: = \: 0.
\ee{MORE-THAN-BJORKEN}
For a one-dimensional cylindrical expansion, $V=\pi R^2 \tau$
with the proportionality constant $dV/d\tau = \pi R^2$, 
the above equation describes Bjorken's flow:
\be
 \frac{de}{d\tau} + \frac{e+p}{\tau} \: = \: 0.
\ee{BJORKEN-FLOW}
If a transverse flow is present, the change of the local volume 
defined by the local four-divergence of the relativistic velocity field
(eq. \ref{FOUR-DIV}) may lead to a more complicated $V(\tau)$ relation.
This effect can only be explored by solving Euler's equation, too.
Some special analytical solutions without comoving pressure gradient,
including transverse flow are given in \cite{BiroAnalHydro}.

In our further analysis we display the cooling, i.e. the evolution of the
temperature  $T$ as a function of the expanding volume $V$. This way
we do not specify the time evolution of the volume factor, or the
four-divergence of the flow; this presentation is independent of the
details of the transverse flow.

\begin{figure}

\includegraphics[width=0.45\textwidth,angle=-90]{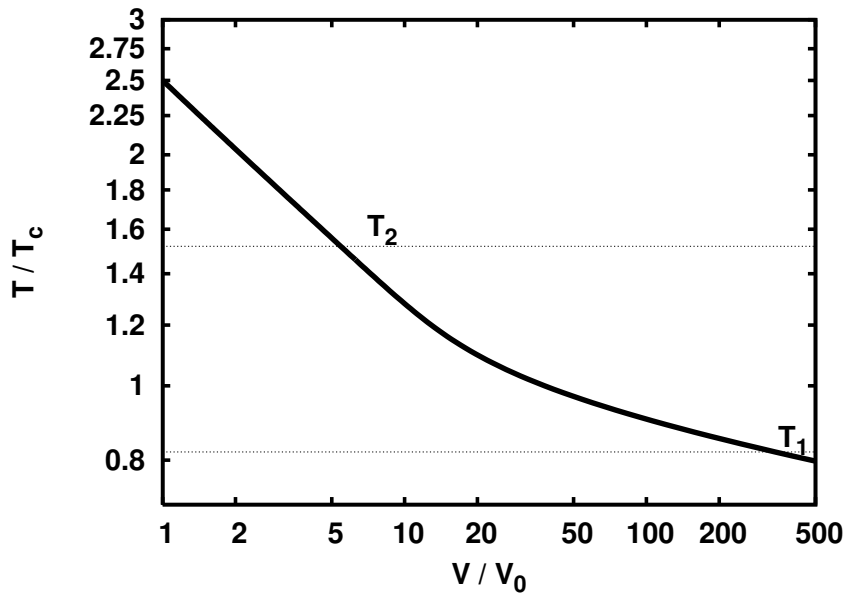}

\caption{\label{FIG2} 
  The adiabatic cooling of QCD matter as a function of the expanding
  volume. The horizontal lines belong to the temperatures $T_2$ and
  $T_1$, respectively.
}
\end{figure}

By using lattice QCD equation of state a first integral of the cooling-expansion
law can be obtained by keeping $S=Vs=V\partial p / \partial T$ constant.
In Fig.\ref{FIG2} the temperature is plotted against the expanding volume
of a matter satisfying an equation of state fitted to lattice QCD
simulations. The starting point was $T=2.5T_c$, the endpoint $T=0.8T_c$.
The change from a fast cooling to a slower one
occurs near $T_c$; the double-logarithmic plot shows a kick in the
power-law behavior clearly.

\begin{figure}

\includegraphics[width=0.45\textwidth,angle=-90]{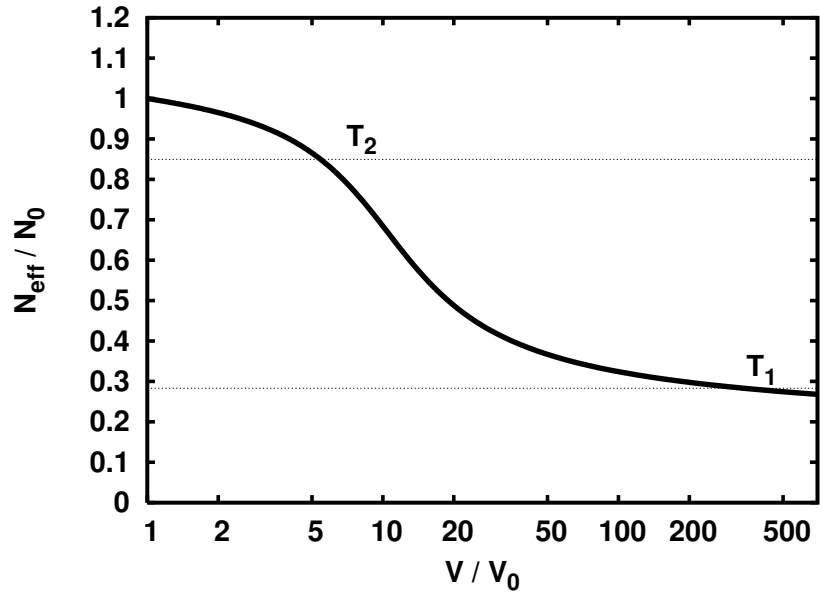}

\caption{\label{FIG3} 
  The ideal number of particles $N_{{\rm eff}}=pV/T$ as a function of the expanding
  volume by adiabatic cooling of the QCD matter. The temperature labels $T_2$ and $T_1$
  indicate the corresponding volumes.
}
\end{figure}

In Fig.\ref{FIG3} the ideal particle number $N_{{\rm eff}}=pV/T$ is plotted as a function
of the expanding volume $V$. Both axes are scaled with the respective initial values.
The steepest drop can be observed around the $T=T_c$ temperature region. 
It is realistic that $N_{{\rm eff}}$ is reduced by almost a factor of three between $T_2$ and $T_1$,
while the volume expands to several ten times of its original size. 
We conclude from this observation that quark-coalescence models 
can be successful. The number of colors is reduced out from the effective
number of degrees of freedom due to confinement even during a mild expansion,
like an ideal fluid.
Further non-equilibrium effects are likely to make the expansion
faster in real experiments, but nowadays ab initio QCD calculations cannot yet
deal with such a situation.

In conclusion we observe that the lattice QCD equation of state leads to
a reduction of about one third in the number of equivalent quasiparticles
between the endpoint temperatures of the linear part in the
$N_{{\rm eff}}/S - T$ curve.
The upper temperature $T_2$ is around the phase transition temperature of pure gauge field systems, 
and the lower temperature $T_1$ is near to the hadrochemical decoupling temperature observed in
relativistic heavy-ion collisions. This may signalize that the hadronization
of quark matter is dominated by recombination processes of massive quarks,
as it is assumed in various coalescence models\cite{ALCOR,COA2,COA3,COA4}. 
This reduction can take place
at constant total entropy, contrary to some claims raised at diverse occasions.


\vspace{7mm}
{\bf Acknowledgment}
This work has been supported by the Hungarian National Research Fund,
OTKA (T49466).


%

\end{document}